\documentclass[10pt]{article}

\usepackage{amsmath}
\usepackage{amssymb}
\usepackage{graphicx}
\usepackage{cite}
\usepackage{color} 
\usepackage{times}
\usepackage{subfig}
\usepackage{array}

\usepackage{caption}

\makeatletter
\renewcommand{\@biblabel}[1]{\quad#1.}
\makeatother

\date{}

\begin{document}

\begin{flushleft}
{\Large
\textbf{Understanding disease control: influence of epidemiological and economic factors.}}
\\
Katarzyna Ole{{\'s}}$^{1,2,\ast}$,
Ewa Gudowska - Nowak$^{1}$,
Adam Kleczkowski$^{2}$
\\
\bf{1} M. Kac Complex Systems Research Center and M. Smoluchowski Institute of Physics, Jagiellonian University, ul. Reymonta 4, 30-059 Krak{\'o}w, Poland 
\\
\bf{2} Department of Computing Science and Mathematics, University of Stirling, Stirling FK9 4LA, United Kingdom
\\
$\ast$ E-mail: kas@cs.stir.ac.uk
\end{flushleft}

\section*{Abstract}
We present a local spread model of disease transmission on a regular network and compare different control options ranging from treating the whole population to local control in a well-defined neighborhood of an infectious individual. Comparison is based on a total cost of epidemic, including cost of palliative treatment of ill individuals and preventive cost aimed at vaccination or culling of susceptible individuals. Disease is characterized by pre-symptomatic phase which makes detection and control difficult. Three general strategies emerge, global preventive treatment, local treatment within a neighborhood of certain size and only palliative treatment with no prevention. The choice between the strategies depends on relative costs of palliative and preventive treatment. The details of the local strategy and in particular the size of the optimal treatment neighborhood weakly depends on disease infectivity but strongly depends on other epidemiological factors. The required extend of prevention is proportional to the size of the infection neighborhood, but this relationship depends on time till detection and time till treatment in a non-nonlinear (power) law. In addition, we show that the optimal size of control neighborhood is highly sensitive to the relative cost, particularly for inefficient detection and control application. These results have important consequences for design of prevention strategies aiming at emerging diseases for which parameters are not known in advance.

\section*{Author Summary}

Designing strategies for disease control is the key goals of epidemiological modeling. Traditionally such strategies have been formed to stop an outbreak as quickly as possible regardless of costs. However, there is a growing appreciation that a successful strategy should minimize the total cost of an epidemic. Thus, it might be more desirable to limit the public health measures or even refrain from treatment. We consider a model of a local disease transmission on lattices. Previous studies have shown existence of an optimal control neighborhood corresponding to a minimal cost. Epidemiological factors: infectiousness, dispersal range, symptoms appearance rate and speed of treatment affect the choice of the optimal control radius. However, a detailed study of such a relationship is still missing, even though in practice approximations have been used to control  epidemics like foot-and-mouth disease or citrus canker. We show that the ring prevention is proportional to the infection neighborhood, but this relationship depends on detection and treatment time in a power law. We believe that our results play crucial role in designing control for an emerging disease with incomplete knowledge of its properties.

\section*{Introduction}
The network-based approaches are a common tool in epidemiological studies \cite{newman2010networks}. These individual-based methodologies allow incorporating the diverse patterns of interaction that underlie disease transmission and have been proved to capture topology of populations \cite{keeling2005models,gastner2006optimal}. An interesting aspect of such studies, with an obvious goal to target spread of the disease, is identification of optimal strategies for the control of a disease under additional constraints \cite{barrett2003global,rowthorn2009optimal,ndeffo2010optimization}.
Network modelling has been successfully used for many systems in order to design control strategies \cite{ferguson2001foot}. However, there are only very few examples involving realistic models and in particular incorporating economic factors. Conversely, bioeconomic models usually ignore the spatial components of the disease spread \cite{klein2007economic,gersovitz2004economical,boccara1994probabilistic}.

In this paper we present a combined epidemiological and economic model to address the problem of optimization of disease control on networks with incomplete knowledge.
Two main sources of costs can be associated with a disease outbreak and control: the palliative cost associated with disease case and costs of measures aimed at preventing further cases.
\cite{kleczkowski2006economic,Kleczkowski2011}. 
The objective of preventive actions is to lower the total cost by investing e.g. in vaccination at the initial stages of the epidemic or culling of infected individuals.

Work so far has shown that optimum control strategy exists but the relationship between the details of it and the model parameters is still elusive.\cite{ferguson2001foot}.
 In our approach, we define a measure of the total cost (the severity index, $X$)  and analyze the influence of the epidemic parameters on its minimization. Our previous research has established existence of optimum control scenarios (Global Strategy (GS), Local Strategy (LS), Null Strategy (NS)). We have presented that the choice of optimal control depends on economic factors but not on epidemiology. This paper fills the gap by performing the sensitivity analysis of epidemiological parameters (such as infected neighborhood size, detection and treatment time) on details of Local Strategy.

Our principal objective is to identify optimal strategies for eradication the disease by determining the threshold size of the control neighborhood. In the proposed model, the neighborhood order $z$ is introduced as a measure of either the distance that the disease can spread (epidemic neighborhood), or the spatial extension of the control measures in a single "event" (control neighborhood). To investigate how limited resources should be balanced between disease detection and eradication, we analyze combined effects of the average time until detection and the treatment rate on optimal control size of the neighborhood. Ability to design control strategy even with limited knowledge make us prepare for emerging diseases.
  
\section*{Model}
We assume that individuals are located at nodes of a regular (square) lattice that
represents geographical distribution of hosts. On this lattice, we define a
local neighborhood of order $z$ as a von Neumann neighborhood in which we include  $z$ shells and $\phi(z)=2z(z+1)$ individuals, excluding the central one. Accordingly,  $z=0$ corresponds to a single individual, which means that this individual is not in contact with anyone, $z=1$ corresponds to 4 nearest neighbors while $z=\infty$ corresponds to the whole population in the limit of infinite size of the system. 

The epidemiological model is a standard SIR (Susceptible-Infected-Removed)
model \cite{anderson1991infectious}, modified to include pre-symptomatic and symptomatic stages of the disease and to account for detection and treatment (cf. fig. 1). All individuals are initially susceptible ({\bf S}) and the epidemic is initiated by introduction of several infected ({\bf I}), pre-symptomatic individuals. Each of infected individuals (symptomatic and pre-symptomatic) stays in contact with a given (fixed) number of other individuals in its infection neighborhood of order $z_{inf}$.
  After infection, the susceptible individual moves first to infected, pre-symptomatic class, ({\bf I}).
compartments. 
It can further infect its neighbors with probability $f$ per a contact but cannot be treated yet.
As symptoms develop with probability $q$, individual moves to {\bf D} class and can be detected. It is still infectious but can spontaneously recover with probability $r$ and accordingly, move to a recovery class, ({\bf R})  and cannot be further infected or treated.

Detection triggers the control process which becomes activated with probability $v$. In consequence, all individuals (except {\bf R}) within control neighborhood of size $z$ centered at the detected host, transfer to the treated class {\bf V}. 
The order of control neighborhood $z$ may be different from the order of infectious neighborhood $z_{inf}$  and is typically larger. Accordingly, the group of individuals subject to the treatment is composed of at least one individual and a mixture of susceptible and infected pre-symptomatic and symptomatic individuals. For convenience, we extend the definition of the neighborhood $z$ to capture situations when  no spatial control is applied ($z=-1$), or when the treatment is applied solely to the detected individual ($z=0$).

Numbers of individuals in each class are denoted by $S$, $I$, $D$, $R$ and $V$, respectively with $N$=\textit{S+I+D+R+V} 
being the total constant number of individuals in the population.

In order to investigate the optimal control strategy, we need to compare value of future benefits (reduction of infection cases) with the value of future and current costs associated with a particular choice of measures in disease control and treatment.
In this paper we allocate the costs to two groups: 
\begin{equation}
X(z, \infty)=R(z,\infty)+ cV(z,\infty).
\label{sev}
\end{equation}
The first term represents the palliative cost and is associated with individuals who never become detected and therefore spontaneously move into the {\bf R} class. The second term describes costs associated with treatment of detected individuals and their neighbors and assumed to be proportional to the number of treated individuals $V$. In the above formula $c$  represents a cost of treatment relative to the cost of  infection and $z$ stands for the control neighborhood size. Both estimates of $R(z,\infty)$ and $V(z,\infty)$ are evaluated at the end of a single simulation run. The optimal strategy is determined by the minimal value of the severity index $X$. The Minimum of $X$ and the corresponding value of $z$ gives the optimal size of control neighborhood, $z_{c}$, see fig. \ref{x2} for illustration.

\subsection*{Simulations}
Monte Carlo simulations have been performed on a regular grid of 200 by 200 cells with periodic boundary conditions. This choice of size has been dictated by a trade off between numerical efficiency and avoidance of small-size effects which could influence results. Additional numerical tests proved the consistency of results for different system sizes.

Epidemics were initiated by addition of $40$ infected individuals to an otherwise susceptible population. The order of infection neighborhood $z_{inf}$  has been varied from $1$ to $8$. Each simulation run has been continued until $I(t) + D(t) = 0$  (i.e. up to the time when no further infection can occur). The severity index $X$ has been evaluated from the formula eq.(\ref{sev}). In the simulation, the minimization of the severity index is achieved by sweeping through different values of the control neighborhood size, $z$. For each value of $z$ only a single simulation has been performed. Based upon a set of $X$ values for different $z$, the actual minimal value of $X$ and the corresponding value of control neighborhood size, $z$ are found. This procedure has been repeated 100 times to yield representative  average values of $z_{c}$ and $X_{c}$ and their corresponding standard deviations.

\section*{Results}
The long time ($t\rightarrow\infty$) behavior of the model in the absence of control (Null Strategy, NS, i.e. $z=-1$) is determined by the probability $f$ of passing the infection to a susceptible node from any of its neighbors within the neighborhood size ranging from 4 ($z=1$) to 144 ($z=8$). For small $f$, the infection quickly dies out. Disease spreads invasively over the population  for large $f$, when no control is applied, $X(z,\infty)\propto R(z,\infty)\simeq N$. When $z\geq 1$, the ratio $R/N$ declines with the order of the control neighborhood. However, at the same time the number of treated individuals $V$ increases contributing to the total cost $X$, cf. eq.(\ref{sev}). 
For $c \neq 0$, $X(z)$ is either a monotonic function of $z$ for small values of $f$ or a non-monotonic function for highly contagious disease (large $f$), see fig. 2.

Three regions can be identified in dependence of $z_c$ on $c$ and $f$ \cite{Kleczkowski2011}, see fig. 3. For small values  of $c$, Global Strategy (GS) is dominating, whereas for large $c$, it is best to refrain from treatment, Null Strategy (NS), fig. 3. 

Although the location of the minimum of $X(z)$ varies with increasing $f$ and $c$ values (see figs. \ref{x2},  \ref{fig.3}), a relatively wide plateau region with an almost constant  $z_c$ develops for intermediate values of $c$ and $f$ and corresponds to the local strategy (LS), fig.3. Since within this parameter domain $z_c$  depends only weakly on $f$ or $c$,  we have further explored dependence of $z_c$ on epidemiological parameters: $z_{inf}$, $q$, $v$ and subsequently analyzed how this dependence varies with $c$.

We first explore dependence of $z_c$ on the size of infection neighborhood for $c=1$, see fig. \ref{zczinf}. The relationship can be accurately approximated by a linear function for a wide range of parameters, infectiousness $f$ (fig.\ref{zczinf}a), the rate at which symptoms appear, $q$ (fig.\ref{zczinf}b) and the treatment rate, $v$ (fig.\ref{zczinf}c) for $z_{inf} \in [1,8]$.

As already seen in fig. 3, infectiousness $f$ hardly affects the slope and intercept of the linear relationship, fig.\ref{zczinf}a.
 Increasing $q$ and $v$ causes the lines to shift towards lower values of $z_c$, with major changes in the intercept but slope only slightly affected, fig.\ref{zczinf}b,c.

However, the relationship between $z_c$ and $q$ (or $v$) for fixed $z_{inf}$ is non-linear. It is more convenient to consider $1/q$ instead of $q$; $\tau_q = 1/q$ also has an interpretation of average time till detection of symptoms. Similarly, $\tau_v = 1/v$ can be interpreted as an average time till treatment.

Broadly speaking, $z_c$ increases with $\tau_q$ and $\tau_v$, fig. 5. This is consistent with the following mechanism. Consider a single infected but pre-symptomatic individual. The disease focus centered on it will spread until appearance of symptoms after time $\tau_q$. Thus, the longer it takes to discover symptoms of the disease, the farther the disease would spread from its original focus. As a consequence, the infected area becomes larger and so does $z_c$. Similarly, the longer time from detection until treatment, the further the disease moves away from original focus. As a result, the control size grows with increasing treatment time.

Intriguingly, it appears that $z_c$ is not linearly related to $\tau_{q}$ (and $\tau_v$) but follows a power law: $z_c=\alpha_q \tau_q^{\beta}$ and $z_c=\alpha_v \tau_v^{\beta'}$ eq.(3) (see fig. 5) with exponents well below 1.

$\beta,\beta'$ are similar for a range of $z_{inf}$, i.e. $\beta \in (0.14, 0.25)$, $\beta' \in (0.10, 0.27)$ for $z_{inf}\in (1,8)$ within the plateau regime of an optimal control radius of the epidemic (fig. 3).

While fig. 5 is representative of results for $c \leq 1$, moving $c$ just beyond $c=1$ causes a dramatic change  in the $z_c(\tau_v)$ dependence for large values of $\tau_q$ and $\tau_v$, corresponding to detection and vaccination time comparable with duration of epidemics (approximately $10^4$ time steps for large values of $\tau_v$ and $\tau_q$). $z_c$ decays abruptly for increasing times $\tau_q$, $\tau_v$, as illustrated in fig. 6. This change is associated with very inefficient control (long time till detection, $\tau_q\gg 1$ and long time from detection to treatment, $\tau_v\gg 1$). If the cost of control is lower or equal to the cost of palliative care, it is still better to treat, even though we are not very efficient with treatment and most patients are spontaneously removed. However, if the cost of vaccination is only marginally higher than the cost of untreated case, prevention is no longer cost-effective. We also note that it is only a combination of very long values of $\tau_q$ \textit{and} $\tau_v$ that leads to a limited rage of application of the scaling formulas ($z_c=\alpha_q \tau_q^\beta$ and $z_c=\alpha_v \tau_v^{\beta'}$).

The scaling region of $z_c$ as a function of $\tau_q$ and $\tau_c$ also depends on $c$ in a fashion reminiscent of fig. 3. For small values of $c$ Global Strategy of treating everybody is optimal regardless of the parameters, cf. fig. 3 with fig. 7. In contrast, Null Strategy is optimal for large $c$ (figs. 3 and 7). The region where Local Strategy is optimal occupies the region near $c=1$, but it becomes narrower when the disease is more infectious (fig. 3) or when the control is less efficient (for increasing values of $\tau_q$ (fig. 7a) and $\tau_v$ (fig. 7b). Within this region, $z_c$ is given by scaling formulas. As seen before, $c=1$ is a special case asymptotically associated with a breakdown of LS for very large or very small $f$ (fig. 3) and very large values of $\tau_q$ and $\tau_v$ (fig. 7). 

\section*{Discussion}

In order to design a successful strategy for controlling a disease we need to take into account not only epidemiological and social factors (including the topology of the social network of contacts and in particular $z_{inf}$), but also economic considerations. Some of these factors might be unknown or hard to estimate, particularly in real time as the epidemic unfolds. It is therefore crucial to understand the relationship between the optimal control strategy and parameters, for a wide range of possible values. It is even more important to establish those processes and parameters to which a selection of optimal strategy is not particularly sensitive, as this allows us to find strategies that can be designed in advance, even without knowing their actual values for a given emerging disease. In our previous paper we have shown that for a given set of $z_{inf}$, $q$ and $v$, the broad choice of the strategy is determined by the relative cost of the treatment, $c$. For small values of $c$, GS is optimal, for large values of $c$, NS. Close to $c=1$, a LS dominates and the detailed value of the control neighborhood $z_c$ depends on the epidemiological parameters, although not on $f$ in a wide range. In this paper we extend this analysis to include other epidemiological parameters. In particular we show that the broad division between GS (for $c\ll 1$), NS (for $c\gg 1$) and LS (for $c\simeq 1$) holds for a wide range of parameters $q$ and $v$ (inverse of time to detection and inverse of time to treatment, respectively), fig. 7.

Three other key results emerge from our analysis. Firstly, it is very important to match scale of control to the scale of infection dispersal. This has already been seen in other papers \cite{Gilligan:2007cv}, but this is the first time we show it for spatial control on networks in the presence of economic evaluation. However, we also show that the size of the control neighborhood is not just simply equal to the size of the infection neighborhood (see fig. 4 and compare the scale of horizontal and vertical axes). In the presence of pre-symptomatic individuals ($\tau_q\gg 0$) and in the face of delays associated with application of control  ($\tau_v\gg 0$) we need to extend $z_c$ well beyond $z_{inf}$. The relationship between $z_{inf}$ and $z_c$ is one of the key formulas for planning response to epidemics. It enables authorities to plan actions aiming at eradication of the disease by setting a sufficiently large -- but not too large -- zone of eradication around each detected case. Traditionally, such recommendations are based on the dispersal patterns of the disease, although increasingly simulation models are used. This procedure has led to establishment of the 1,900ft rule for citrus canker \cite{Gottwald2001} whereby all citrus trees are cut down within this radius from every affected tree and the 3km/10km rule for foot-and-mouth disease \cite{defra}. 

However, our results show that the relationship between $z_c$ and $z_{inf}$ is non-trivial and in particular it involves non-linear functions of $\tau_q$ and $\tau_v$. Although we are still far from being able to provide a formula relating $z_c$ to all epidemiological parameters, our result stresses importance of using models to design control strategies \cite{Kao2002}. 

Finally, we show that $c=1$ is a special case. In particular, we show high sensitivity of $z_c$ to changes in $c$ for large values of $\tau_q$ and $\tau_v$. Thus, if the symptom detection time ($\tau_q$) and reaction time ($\tau_v$) are both long, small change in $c$ leads to very big changes in $z_c$, see fig. 6 and 7. Without knowing the exact value of $c$ it is therefore very difficult to design the strategy in this case. Suppose we believe that $c>1$ and therefore we chose a small value of $z_c$ based upon fig. 6b. However, if in reality $c\leq 1$ (although very close to 1), $z_c$ should be close to 50 (fig. 6a). This shows the importance of knowing what the actual value of $c$ is \cite{Kleczkowski2011} estimated that for vaccination $c=0.01$-0.85, but  can be larger than 1 for culling. 

Our studies can be extended in several ways. The current work assumes relatively short overall time length of each epidemic is very short and so no discounting is applied when the costs and benefits are estimated. We also assumed that the strategy is unchanged throughout the epidemic and that the network structure is static and very simple. Each of these assumptions can be relaxed. Discounting is often used in economics, but we expect for it to have a small impact on our results. Adapting the strategy to the current status of the epidemic often leads to a bang-bang solution \cite{Forster:2007pf}, similar to our distinction between NS and GS. Finally, a lot of attention have been recently given to non-local and random networks (small-world or scale-free networks, \cite{Kleczkowski2011}, \cite{Dybiec:2004fq} and to dynamic networks \cite{Vernon2009} as well as networks with random parameters \cite{Taraskin:2005bq}. Work on extension of our model to account for these heterogeneities is in progress.

\section*{Acknowledgments}
We are very grateful to Bartek Dybiec for useful discussions.

\section*{Author Contribution}
Proposed a model: AK KO.
Performed simulations: KO.
Analyzed the data: KO EGN AK.
Wrote the paper: KO EGN AK.

\bibliography{biblio}

\section*{Figure Legends}
\begin{figure}[!ht]
\begin{center}
\includegraphics[width=4in]{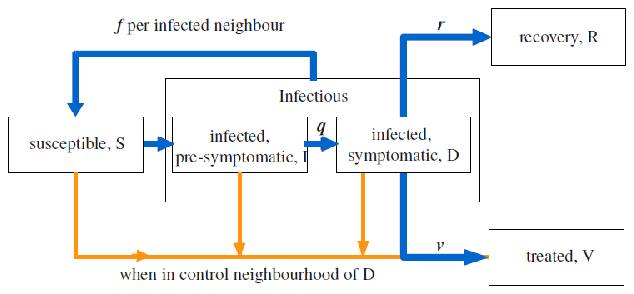}
\end{center}
\caption{{\bf Block diagram illustrating transitions in the model:}
transitions performed at each time step (blue solid lines) and transitions triggered by treatment (orange thin lines).
}
\label{fig.1}
\end{figure}

\begin{figure}[!ht]
\begin{center}
\includegraphics[angle=270,width=4in]{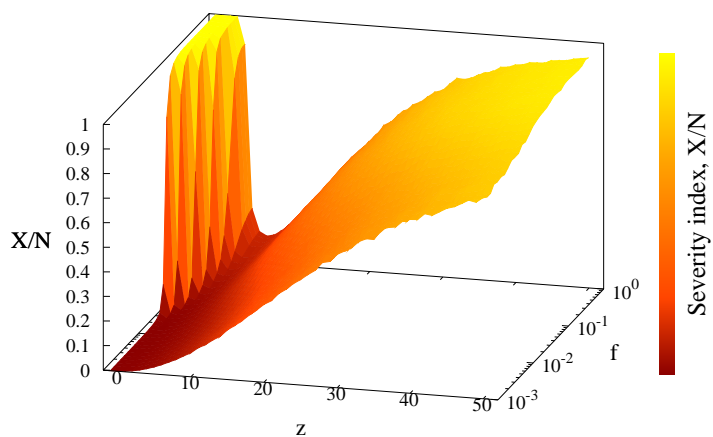}
\end{center}
\caption{{\bf Severity index, X, as a function of the infection rate per contact $f$ and the control neighborhood size $z$.} Simulation parameters: $q=0.5$, $v, r=0.1$ with 40 initial foci and infected neighborhood size set to  $z_{inf}=1$, cost $c=1$.
} 
\label{x2}
\end{figure}

\begin{figure}[!ht]
\begin{center}
\includegraphics[angle=270,width=4in]{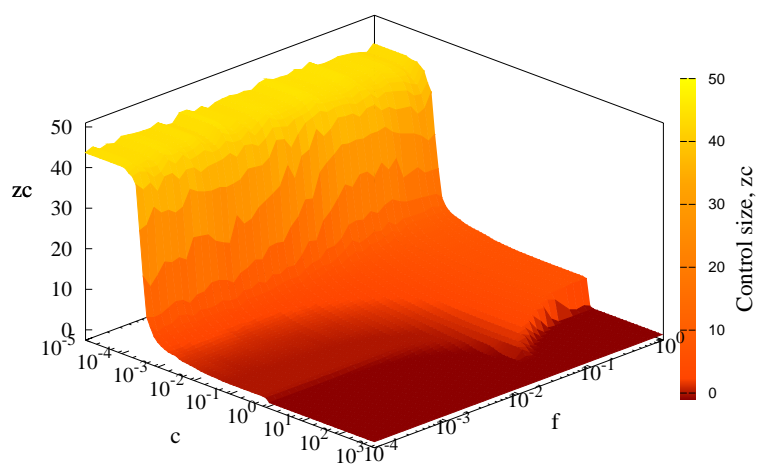}
\end{center}
\caption{{\bf Control neighborhood size as a function of treatment cost $c$ and infectiousness of the disease $f$. }Simulation parameters: $q=0.5$, $v, r=0.1$, with 40 initial foci and $z_{inf}=1$. Control size $z_c>0$ represents local strategy (LS), $z_c=0$ corresponds to the strategy when only the detected  individual is treated and $z_c\geq 30$ denotes GS (more than 99\% of individuals are treated). Null strategy corresponds to $z_c=-1$.}
\label{fig.3}
\end{figure}

\begin{figure}[!ht]
\begin{center}
\includegraphics[width=3in]{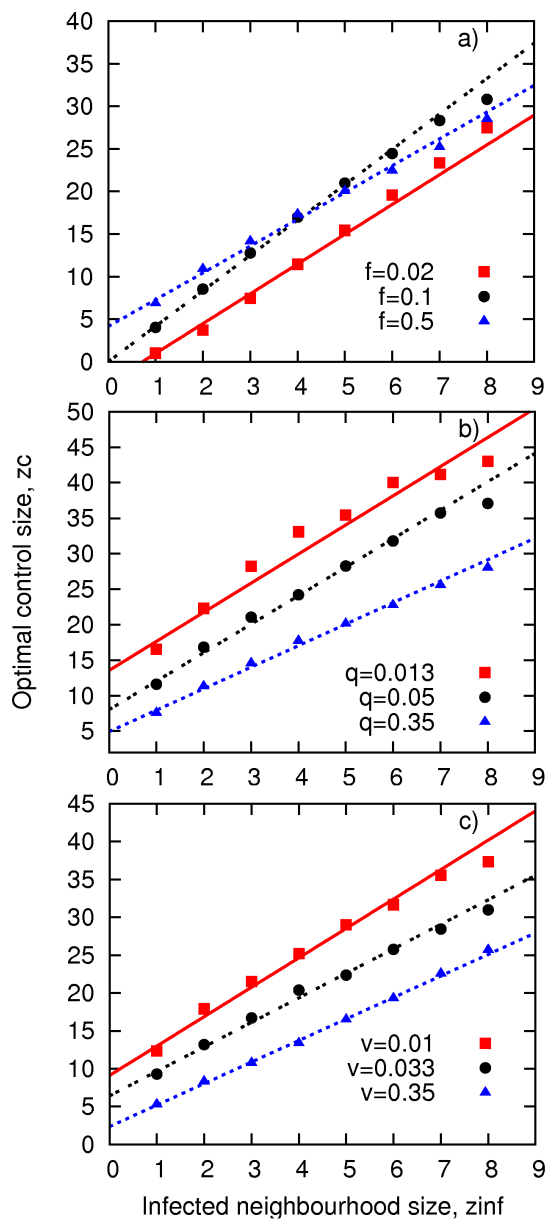}
\end{center}
\caption{{\bf Relationship between $z_{c}$ and $z_{inf}$ for treatment cost $c=1$.} Points mark the simulation results whereas lines correspond to fitted linear function  $z_c = z_{inf}*a+b$. From top to bottom, the following sets of constant kinetic parameters have been assumed: (a) $q=0.5, v=0.1$, (b) $v=0.1, f=1$, (c) $q=0.5,f=1$. 
Errors (standard deviation from the mean) are too small to be visible.}
\label{zczinf}
\end{figure}

\begin{figure}[!ht]
\begin{center}
\includegraphics[angle=270,width=4in]{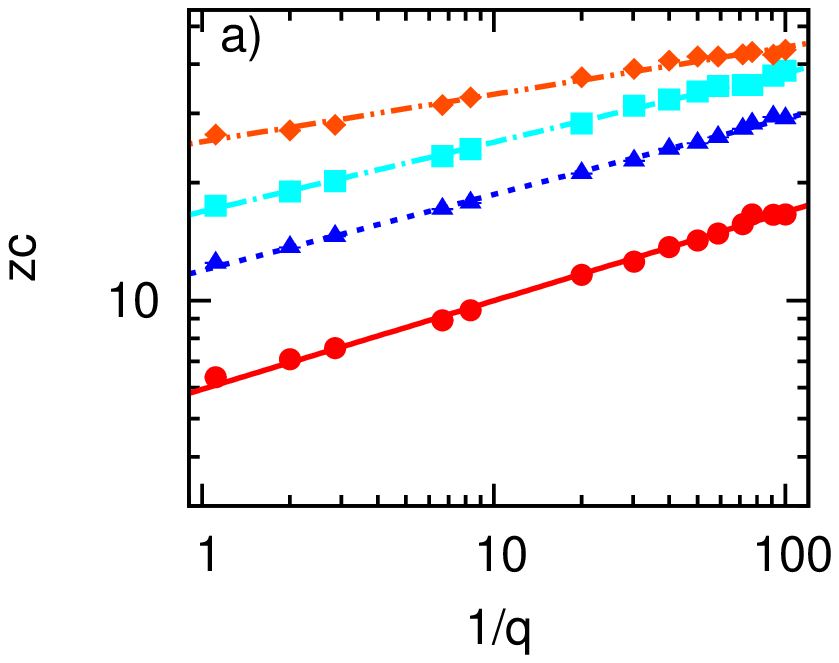}
\includegraphics[angle=270,width=4in]{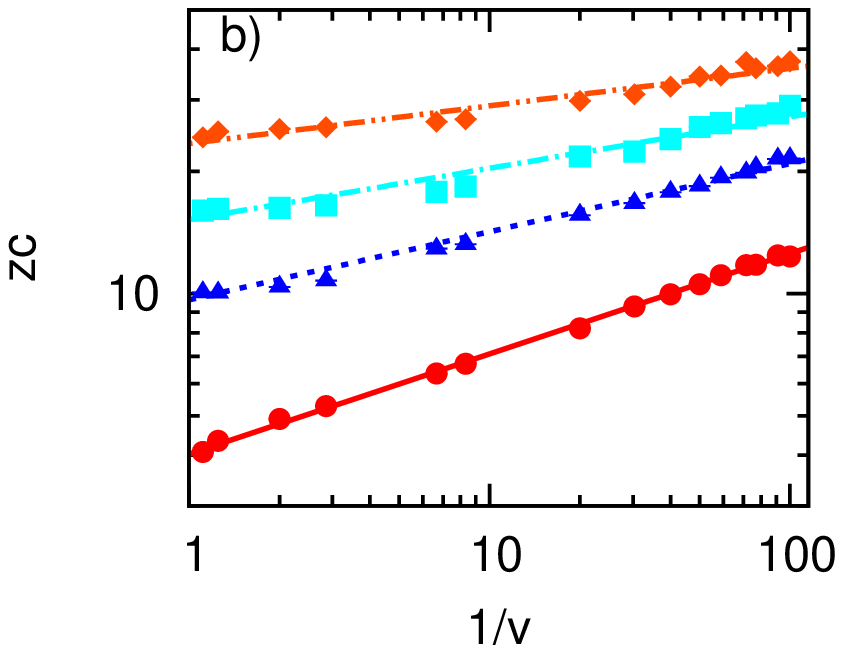}
\end{center}
\caption{{\bf Relationship between $z_{c}$ and $\tau_{q}$ in a) and $\tau_v$ in b).} Points mark the simulation results and lines correspond to fitted functions: 
a): $z_c(\tau_q) = \alpha_q \tau_q ^\beta$ and 
b): $z_c( \tau_v) = \alpha_v \tau_v ^{\beta'}$ for 
red: $z_{inf} = 1$ , navy blue: $z_{inf} = 3$ , blue: $z_{inf} = 5$ , orange: $z_{inf} = 8$.
}  
\label{zc1q}
\end{figure}

\begin{figure}[!ht]
\begin{center}
\includegraphics[angle=270,width=4in]{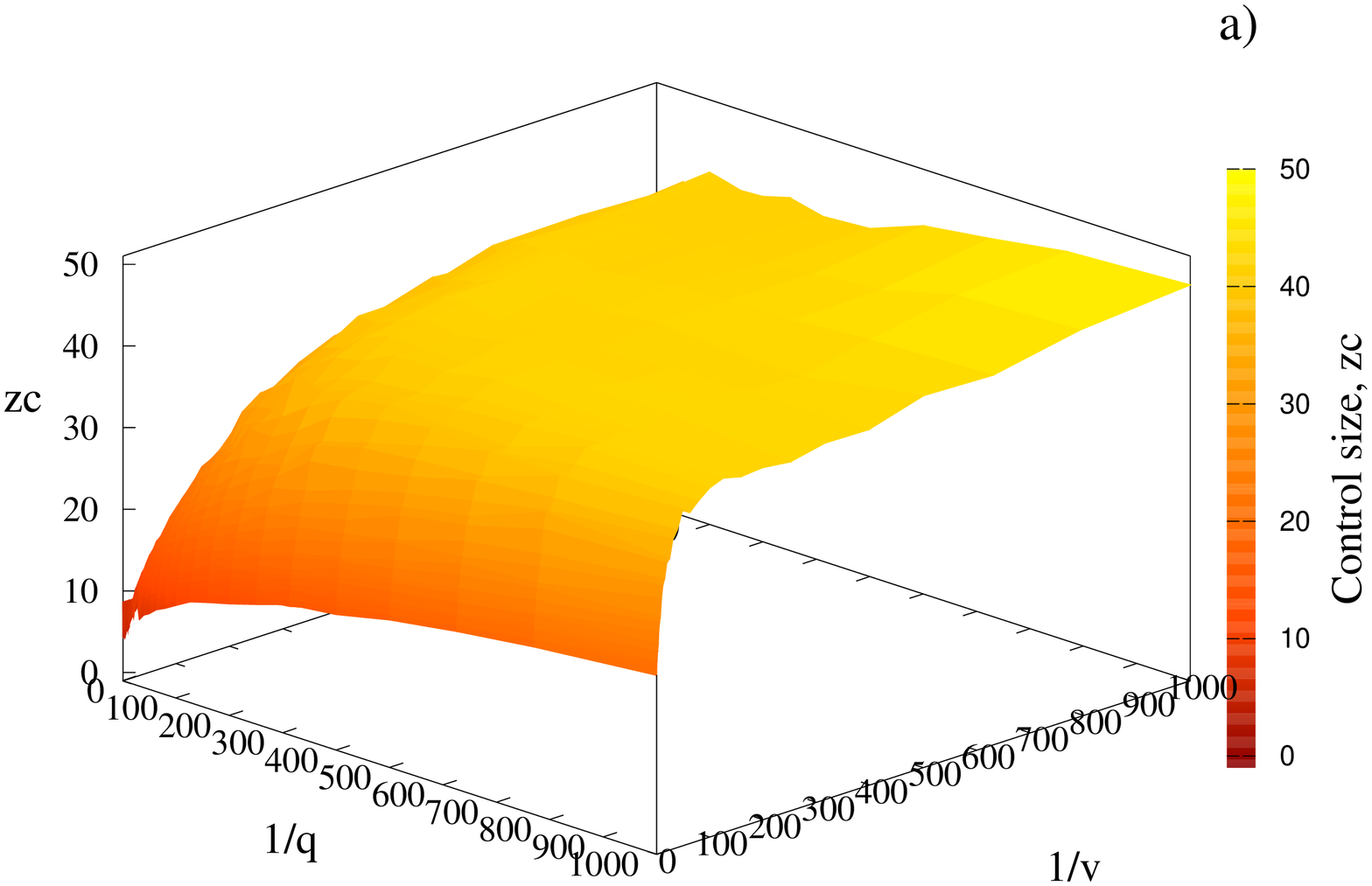}
\includegraphics[angle=270,width=4in]{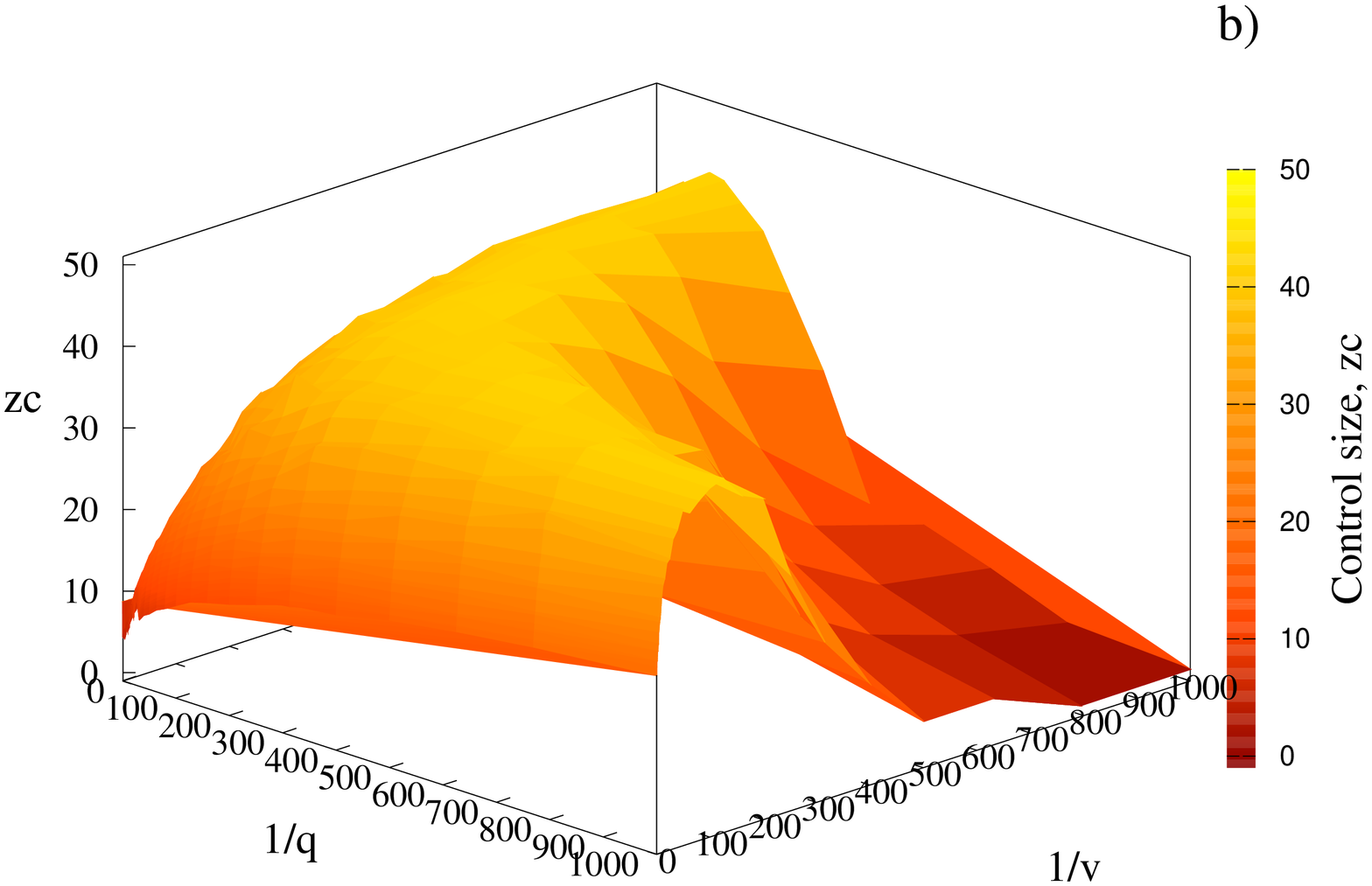}
\end{center}
\caption{{\bf Control neighborhood size as a function of both detection time, $\tau_q$, and recovery time, $\tau_v$ for $c = 1$ in a) and $c=1.001$ in b).}
 Simulation parameters: $f=0.1$, $r=0.1$, $z_{inf}=1$, 40 initial foci. 
 }
\label{mapqv}
\end{figure}

\begin{figure}[!ht]
\begin{center}
\includegraphics[angle=270,width=4in]{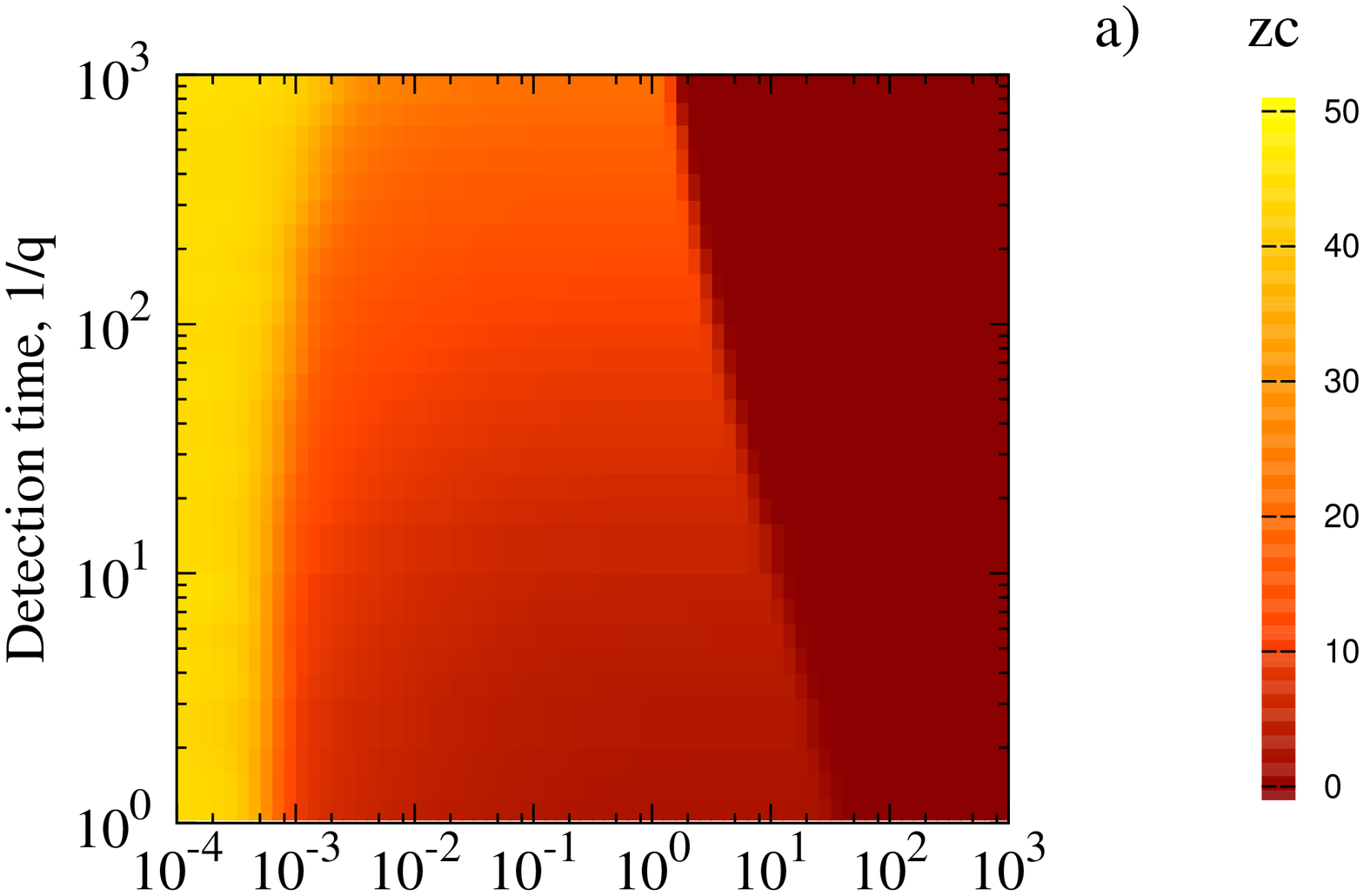}
\includegraphics[angle=270,width=4in]{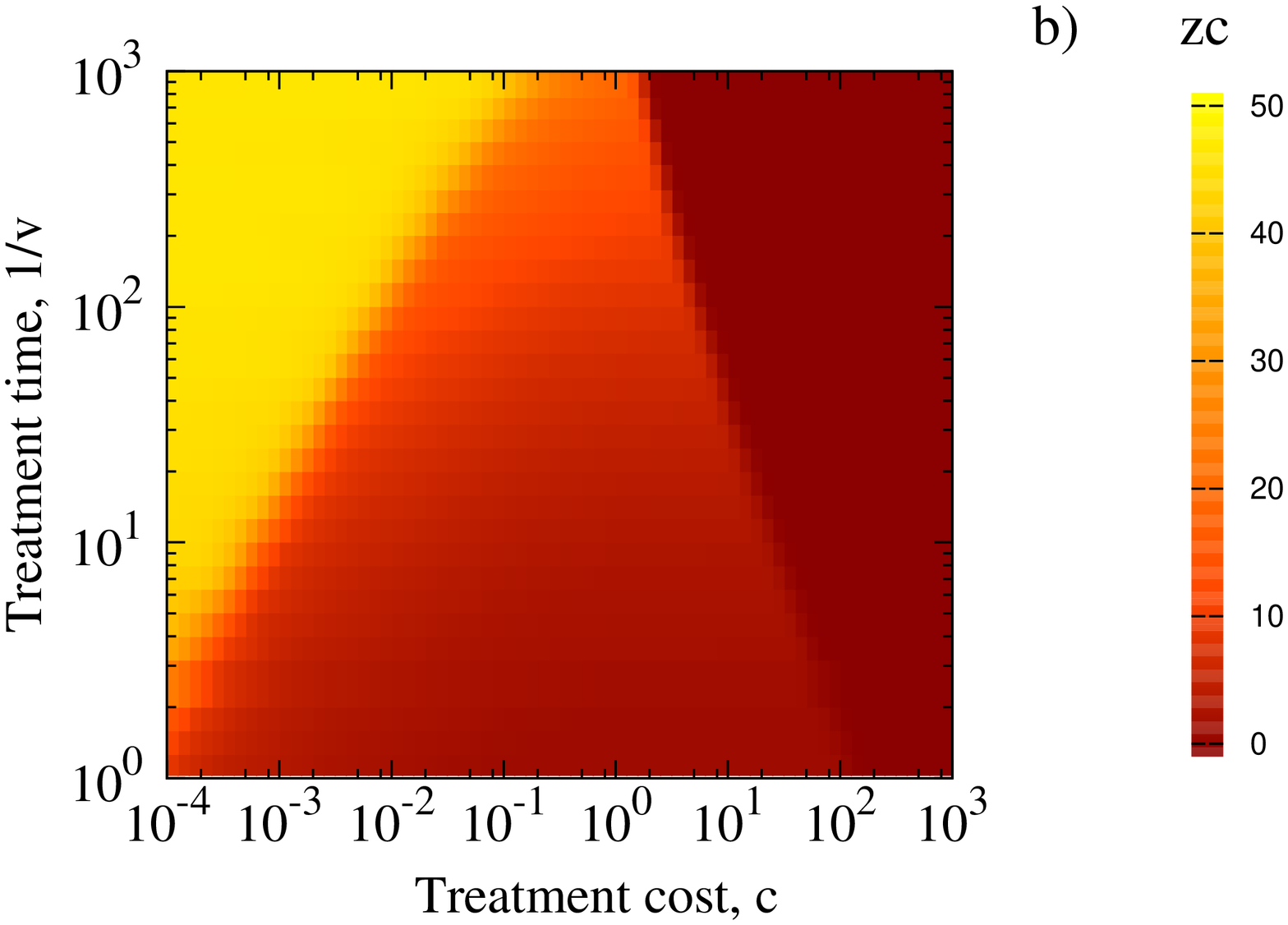}
\end{center}
\caption{{\bf Control neighborhood size as a function of treatment cost $c$ and detection time $\tau_q$ (a) and treatment time $\tau_v$ (b).} Simulation parameters: $f=1$, $q=0.5$,$v, r=0.1$, with 40 initial foci and $z_{inf}=1$. Color borderlines between different regions indicate transition regions among various optimal strategies.}
\label{fig.10a}
\end{figure}

\end{document}